\documentclass[structabstract]{aa}  

\usepackage{graphicx}
\usepackage{txfonts}
\usepackage{natbib}
\begin{document}

\title{Horizontal photospheric flows trigger a filament eruption}
\author{T. Roudier\inst{1}
\and
       B. Schmieder\inst{2}
       \and
       B. Filippov \inst{3}
        \and
        R. Chandra\inst{4}
        \and
        J.M. Malherbe\inst{2}
        }
\offprints{Th. Roudier,\\
\email{thierry.roudier@irap.omp.eu}}
\institute
    {
      Institut de Recherche en Astrophysique et Plan\'etologie, Universit\'e de Toulouse, CNRS, UPS,
      CNES 14 avenue Edouard Belin, 31400 Toulouse, France
\and
Observatoire de Paris, LESIA, 5 place Janssen, 92195 Meudon, France, PSL Research University, CNRS, Sorbonne Universit\'es, UPMC Univ. Paris 06, Univ. Paris Diderot, Sorbonne Paris Cit\'e 
\and
Pushkov Institute of Terrestrial Magnetism, Ionosphere and Radio Wave Propagation of the Russian Academy of Sciences
(IZMIRAN), Troitsk, Moscow 108840 , Russia
\and
Department of Physics, DSB Campus, Kumaun University, Nainital – 263 001, India
}

\date{Received date / Accepted date }
\titlerunning{Horizontal photospheric flows}
\authorrunning{Roudier et al.}

\abstract
{  A  large filament  composed principally of two sections erupted sequentially  in the southern hemisphere on January 26 2016. The central,  thick  part of the northern section   was  first  lifted up and lead to the  eruption of the full filament. This event was  observed in H$\alpha$  with the Global Oscillation Network Group (GONG) and Christian Latouche IMageur Solaire (CLIMSO), and in ultraviolet (UV)   with the  Atmospheric Imaging Assembly (AIA) imager  on board  the Solar Dynamic Observatory  (SDO). }
{The aim of the paper is to relate the photospheric motions below the filament and its environment  to the eruption of the filament. }
{An analysis of the photospheric motions   using Solar Dynamic Observatory Helioseismic and Magnetic Imager (SDO/HMI) continuum images  with the new version of the coherent structure tracking (CST) algorithm  developed to track granules,  as well as large-scale photospheric flows, has been performed.  Following velocity vectors, corks migrate towards converging areas.}
{The supergranule pattern is clearly visible outside  the filament channel but  difficult to  detect inside because the modulus of the vector velocity is reduced in the  filament channel, mainly  in  the magnetized areas. The horizontal photospheric flows are strong  on the west side of the filament channel and oriented towards the filament. The ends of the filament sections are found  in  areas of concentration of corks.    Whirled flows are found locally around the feet. } 
 {  The  strong horizontal flows  with an  opposite direction to the differential rotation create   strong shear  and convergence along the  magnetic polarity inversion line (PIL) in the filament channel. The filament has been destabilized by the converging flows,  which  initiate  an  ascent of the middle section of the filament until the filament reaches the critical height of the torus instability  inducing, consequently, the eruption. The $n$ decay index indicated an altitude of 60 Mm for the critical height.  It is conjectured that the convergence along the PIL is due to the large-scale size cells of convection that transport the magnetic field to their borders.}

\keywords{Sun: filaments, dynamics, photospheric motions}

\maketitle

\section{Introduction}

The physical conditions leading to filament eruptions and coronal mass ejections (CMEs) have been recently reviewed by \citet{Schmieder2013}.   They are based on the existence of flux ropes in the corona  submitted to increasing   electric currents.   The  decrease of  the  magnetic tension that restrains the flux rope favors its eruption.  Twist motions, shears, and canceling flux are observed or involved in the magnetohydrodynamics MHD models  as producing  instabilities of the filament flux rope.   Commonly it is found that eruptions are due to 
converging flows and canceling polarities along the polarity inversion lines (PIL). A filament consists of  a magnetic structure aligned along the PIL with  untwisted magnetic field lines anchored at both ends in opposite magnetic photospheric polarities of  the network as  linear force-free field extrapolations suggest \citep{Aulanier98,Aulanier2002}.  The cool  filament  material is suspended in the dips.  Along an H$\alpha$ filament or prominence,  footpoints or legs are observed with an equidistant distance of about 30Mm, which  we also call barbs when observed on the disk.  A filament  barb is directly related to a parasitic polarity   close to the PIL \citep{Aulanier98,Martin94}. If the parasitic polarity is canceled by merging with   opposite sign polarities, the  barb disappears and the filament is no longer anchored in  place.  Material may be  falling  along the field lines, which   transforms the dips to loops  as was suggested in an MHD simulation \citep{Schmieder06}.    This explains  the counter streaming often observed in rising filaments before eruptions \citep{Zirker98,Schmieder08}. 

In  MHD models, it is shown that canceling flux, twist, or rotation of sunspots induce a strong shear along the PIL. Progressively a flux rope is formed by reconnection of low magnetic field lines in a region of convergence \citep{vanBallegooijen1989,Aulanier2010}. The reconnection is driven by the diffusion of the photospheric magnetic field.  It  allows the flux rope to rise progressively  in the upper atmosphere. When the flux rope reaches  a critical height, a torus-type instability starts accelerating the rise of the flux rope and leading to eruption \citep{Tend1978,Kliem2006}.  The critical altitude is estimated by the computation of the decay index ($n= - dlnB/dlnz$) , which represents the decay of the background magnetic field in the vertical direction. The critical height is defined as the height where the decay index  reaches a critical value slightly dependent on the model of the flux rope used. For a thin straight current channel the critical value of the decay index is equal to unit \citep{Tend1978,Filippov2001}.

\citet{Torok2007} found that for a thin circular current channel a torus instability of the flux rope is initiated when $n$ is larger than 1.5. From an observational point of view, \citet{Filippov2001} and \citet{Filippov2008} performed  a statistical study of quiescent prominences and concluded that
prominences were more prone to erupt when they approach a height where the decay index of the external  field was close to 1. \citet{Zuccarello2016} computed the decay index for different MHD configuration models and derived a possible range for the critical decay index between 1.3 and 1.5 when the apex of the flux rope is considered, otherwise it can be lower.  The difference referred to in these works between the values of the critical index 1 and 1.4 can be considered  unimportant, while the decay index in regions with filaments changes from $\ll$ 1 to 3. The exact value of the critical decay index depends on the shape of the flux-rope axis and the properties of the  flux-rope cross-section as was shown by \cite{Aula2010}. In the model simulations of  Zuccarello et al. (2016), the axis of the flux rope is more curved, which is more appropriate for three-dimensional (3D) modeling. However, even in this model the topmost part of the distribution of magnetic dips, where filament material sits,  is located at n = 1.1 when the instability starts. Since in the works of Filippov and colleagues the relationship between  decay indexes and filament top heights was studied, there is no discrepancy with the results of  Zuccarello et al. (2016).

Surface motion on long-term as well on  short-term scales  is   important for the formation and the stability of filaments.   Photospheric motions are  due to the coupling of the  convection with the  diffusion of the magnetic field at the solar surface. 
 It is important to be able to quantify the  horizontal flows in the photosphere and  follow their evolution.

A  few analyses of surface motions have been done with high cadence and high spatial resolution. Mainly the local correlation tracking (LCT) method is applied to magnetic field polarities  in order to explain eruptions or flares with Michelson Doppler Imager (MDI) data, with a spatial resolution of 1.96 arc sec, and only  recently with Helioseismic and Magnetic Imager (HMI)  data, with a resolution of 0.5 arc sec \citep{Ahn10,Zhou06,Green11,Liu12}.   The techniques of correlation tracking (LCT) \citep{November88,Chae08} have been developed in two ways:  either by tracing the surface flows with the differential affine velocity estimator for vector magnetograms (DAVE4VM) \citep{Schuck08}  or by tracking coherent structures  (CST) on various scales (spatial and temporal) \citep{Roud2012}. 

\citet{Roudier08}  studied the flow pattern in a filament  channel in a bipolar region using LCT,
from MDI magnetograms and Dopplergrams supergranular flow pattern.  It was found that the flow field changes significantly during the eruption phase, measuring an increase of the shear below the point where the eruption starts and a decrease after it \citep{Roudier08}. They found  a pattern in the large scale horizontal flows at the solar surface that interacts with the differential rotation. The local photospheric flows were also measured with a higher spatial resolution (0.5 arc sec) in the filament channel during its eruption phase using  Transition Region and Coronal Explorer (TRACE) 1600 \AA\  to show the coupling between convection and magnetic field  \citep{Rondi07}. Apparently along the PIL both parasitic and normal polarities were continuously swept  by the diverging supergranular flow and canceled, which leads to the filament eruption. 
\citet{Schmieder2014} computed the proper horizontal flows in a filament channel  using the CST method  and conjectured that the shear  flows were responsible for the   counterstreaming along the filament axis and finally for the eruption. Flux cancellation and magnetic shear were already proposed  to play a major role in the filament eruption based on observations \citep{Martin1986,Litvinenko1999,Hermans1986,Martin1998} and models \citep{Priest1987,vanBallegooijen1989,Priest1994}.   Still now, similar measurements are done using  magnetograms showing  the disappearance on a single bipole  by flux cancellation as responsible for a filament eruption \citep{Wang2013,Yardley2016}.   The action of large-scale flows  on the filament (formation and eruption) has not  yet been quantified. In the present study, we take  advantage of the CST method applied to the full Sun data (SDO/HMI) to get flow fields over the whole of the Sun's surface at  high and low  spatial and temporal resolutions. In particular the CST method allows us to get large-scale flows such as solar differential rotation, meridian circulation, and supergranulation flows \citep{Chian2014,Roudier2018}. Previous works use LCT  and many
other methods for detecting approaching bipoles (see the benchmark paper of \citet{Welsch2007}), but no detection of large-scale inflow or shear has
been done. Therefore, the measurement of large-scale flow such as differential rotation or supergranular flow is fundamental with regards to this question of
filament eruption.

We propose in this paper to correlate the horizontal photospheric flows, the H$\alpha$ (see Movie 1), and the UV dynamics (see Movies 2 and 3) of the filament before its eruption. In Sect. 2 we describe the observations of the filament and its dynamics observed with   space- and ground-based instruments. We pay particular attention to the ends or anchorages of the filament  in the network and the progressive lift up of the main body of the filament until it reaches the critical height for developing the torus instability. 
In Sect. 3    the  new CST  method that allows us to   compute  the horizontal photospheric flows  in the filament channel and its environment  is described. The results  are presented in Sect. 4 and discussed in the conclusion from the perspective of the coupling between magnetic field and convection. (Sect. 5).

 \begin{figure*} 
 \centerline{
 \includegraphics[width=0.9\textwidth,clip=]{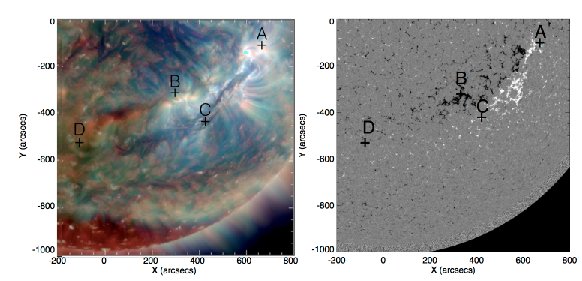}}
 \caption{Huge filament   starting to lift up observed on January 26 2016  at 16:30 UT with SDO/AIA (before the eruption; left panel): combined image of three filters (211 \AA, 193 \AA, 171 \AA; right panel)   HMI  longitudinal magnetogram  showing the magnetic channel of the filament. A, B, C, D are the places  in the network where the magnetic field lines supporting the filament   are anchored. The filament consists of  two sections  or  two filaments  (Filament one: AB and  Filament two: CD). The temporal evolution of the left panel showing the huge filament eruption is available as a movie online. }
 \label{SDO}
 \end{figure*}
 
 \section{Eruption of the filament  }
 \subsection{Overview }
 A long filament  elongated over more than one solar radius (500 arc sec, between latitude $-10\degr$ to $-40\degr$)  was located in the southern part of the  NOAA  AR12486 active region.  This filament has   three different sections: a narrow filament inside the active region between strong polarities on both sides of the PIL, a very wide  part in the middle (300 Mm), and again relatively narrow  patches in the east. In fact the two first sections belong to one filament (Filament one) and the third section to a second filament (Filament two).
  These filaments have  been observed in multi-wavelengths by the  Atmospheric Imaging Assembly (AIA), \citep{Lemen2012} aboard the  SDO satellite (SDO, \citet{Pesnell2012}).  The spatial resolution of  AIA is   0.6 arc sec per pixel, and the cadence 12 s. From the ground the filament was followed by GONG with a cadence of one minute and a pixel of 1 arc sec, and by  CLIMSO  with a cadence of 1 minute and a pixel of 1.2 arc sec (see Movie 1).
 In addition, the magnetic field and the continuum  were continuously observed with the HMI, \citet{Scherrer2012,Schou2012}).  The evolution and the  eruption of this long filament  have been reported by \citet{Zheng2017} and \citet{Luna2017}.
  On January 26 around 17:00 UT, the middle part of the filament started to rise (Figure \ref{SDO}). Long loops in Filament one  with untwisted  threads  representing  magnetic field lines are observed  in AIA 211 -171-193 \AA. These loops 
  are anchored  in positive magnetic polarities of plage regions  that we designate as 'A' in the top right of the Figure \ref{SDO} . The second end  is located   in negative magnetic polarities that we called 'B', far away from the PIL.

  \begin{figure} 
  \centerline{\includegraphics[width=0.5\textwidth,clip=]{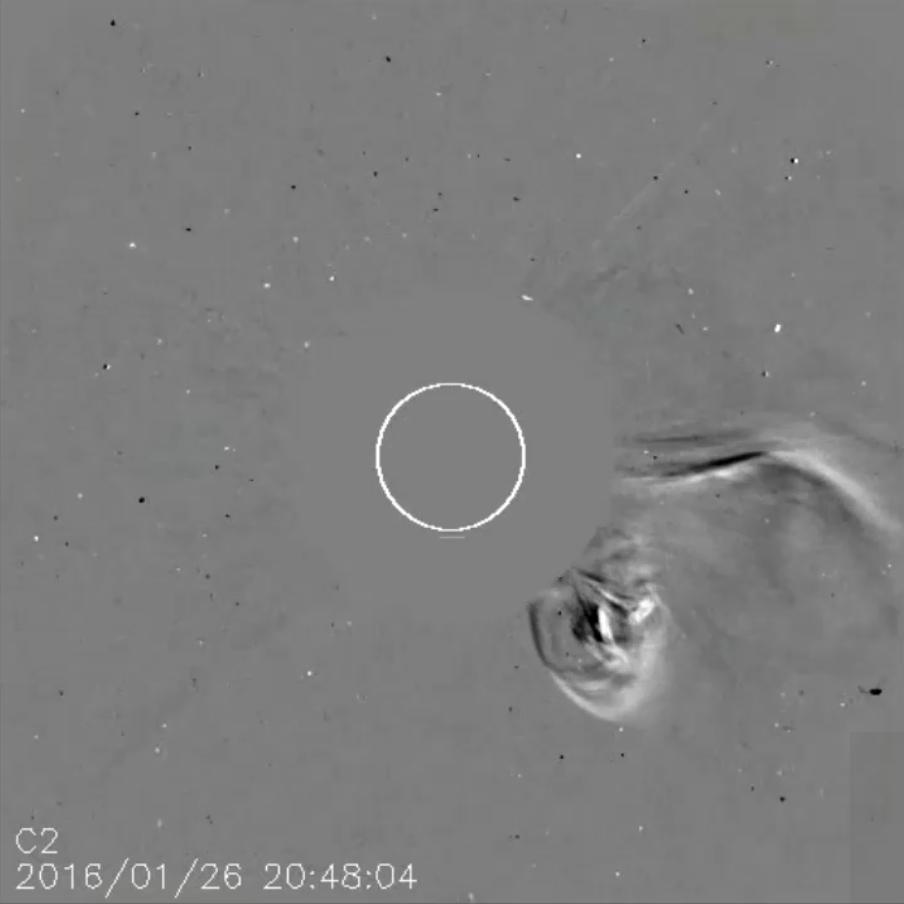}}
 \caption{Filament eruption is related to two CMEs observed by the Large Angle and Spectrometric Coronagraph Experiment (LASCO,/SOHO) coronagraph. The temporal evolution is available as a movie online.}
 \label{C2}
 \end{figure}

  The southern patched section of the filament  (Filament two)  appears to be anchored below the central  thick part  of the filament in C and its other end, D,  is in the  far east. The  section AB (Filament one)  is lifted  up and it erupted at 17:13 UT with untwisted loops  leading to the eruption of Filament two as well (Fig. \ref{SDO}, left panel).
  Material returned and fell in the eastern part of the filament. The loops anchored in B slipped along the magnetic field area  towards the south-east during the eruption allowing the two filaments to merge (see  Movies 2 and 3 attached to Fig. \ref{SDO}). The   material went back and forth exhibiting an oscillation motion   for several hours with a period of 60 min \citep{Luna2017}.  The magnetic structure of this filament has been modeled using the flux inserting method and the above description  of the different anchorages of each filament end is confirmed (see the Fig. 16  in \citet{Luna2017}. A flare is  observed around 1700 UT close to the filament eruption \citep{Luna2017}. This eruption is related to  the two Coronal Mass ejections  (CMEs) observed by  the Large Angle and Spectrometric Coronagraph Experiment (LASCO) coronagraph (see Fig. \ref{C2} and Movie 4).

 \begin{figure*} 
 \centerline{\includegraphics[width=0.8\textwidth,clip=]{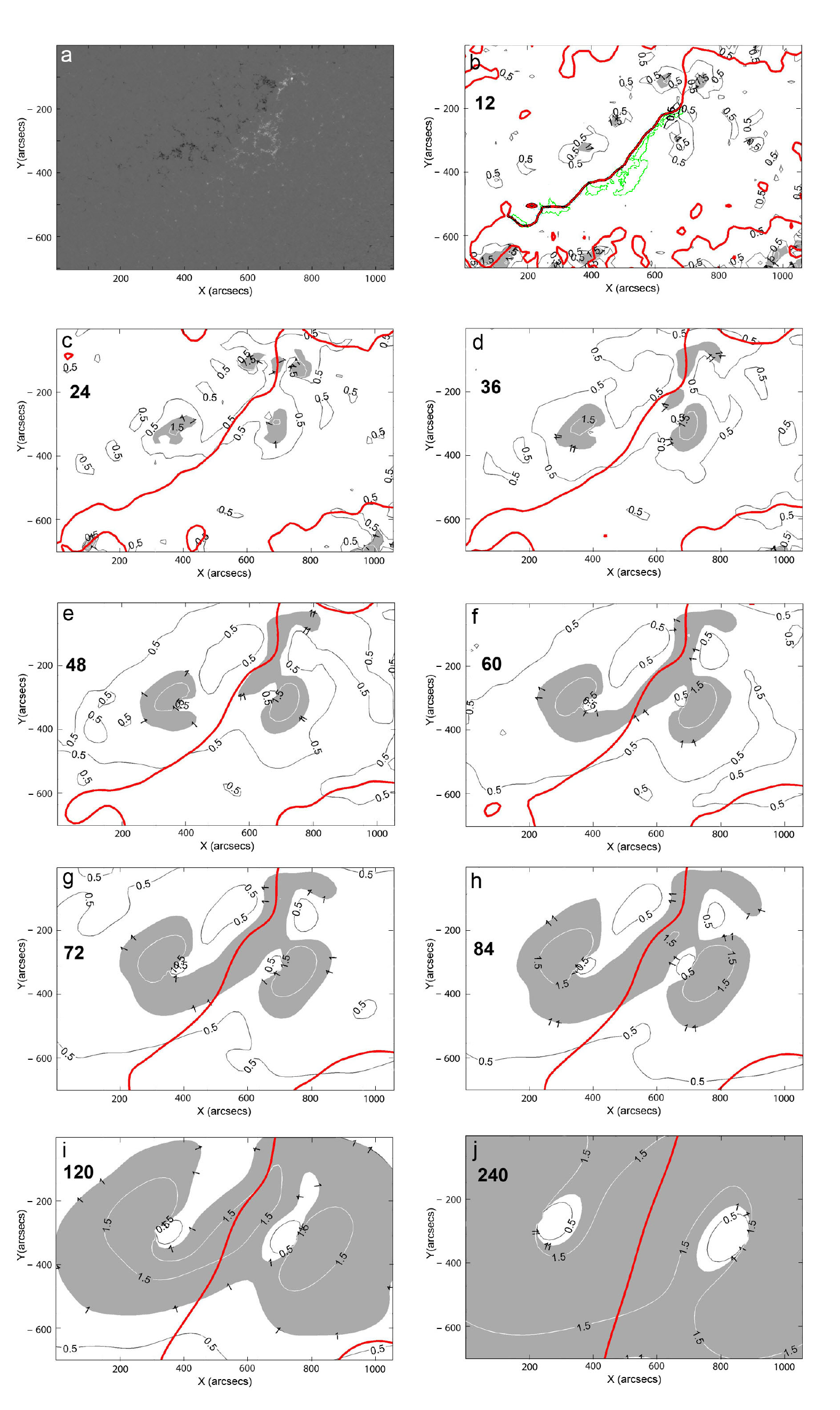}}
  \caption{(a) HMI magnetogram of the selected area taken on  January 26 2016  at 16:30 UT. (b) to (j) panels:  Distribution of the decay index $n$ of the potential magnetic field at different heights above the area shown in first figure  panel (a). Areas where $n > $1 are tinted with gray, while regions with $n <  1$ are white. Red lines indicate the position of the polarity inversion lines (PIL). The green contour shows the position of the filament taken from the co-aligned BBSO H$\alpha$ filtergram. The torus instability  in the zone of the PIL occurs   for decay-index values  $n >1$ .}
 \label{torus1}
 \end{figure*}

 \begin{figure*} 
 \centerline{\includegraphics[width=0.9\textwidth,clip=]{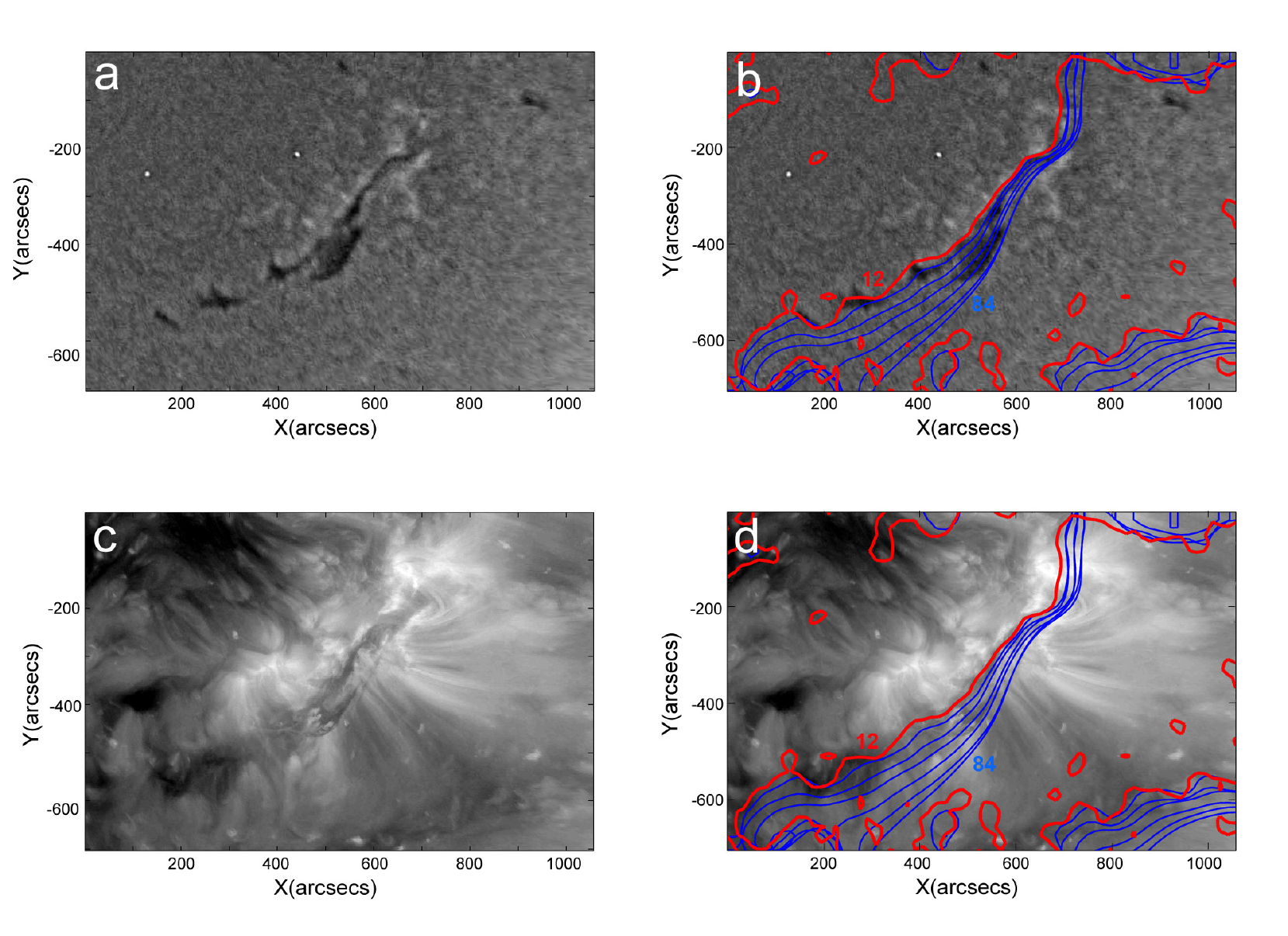}}
 \caption{ Prominence and polarity inversion lines. The top left panel shows the BBSO H$\alpha$ filtergram, the bottom left panel shows the AIA 193 \AA  image of the prominence. The field of view is the same as in Fig. 3a. The panels on the right show the corresponding images with the polarity inversion lines (PIL) at different heights (12 Mm to 84
   Mm) overplotted. The temporal evolution of the filament eruption as seen in the (extreme) UV by AIA in
   1600 \AA, 304 \AA, 171 \AA, and 131 \AA\  is available as a movie online. }
 \label{torus2}
 \end{figure*}

\subsection{Torus instability} 
  
The traditional (logical)  cause of the eruption of this filament is  a   torus instability or a critical loss of equilibrium (nearly the same). The catastrophic loss of equilibrium and the torus instability are  different formulations of the onset of solar eruptions. However, \citet{Aula2010} and
  \citet{Kliem2014} showed  their equivalency.  They are based on the same force balance for equilibrium and produce an onset of eruption at the same point.
This is consistent with  the estimations of the filament height and the decay-index
distribution as we show in the following. 

The decay index characterizes the scale of the magnetic field. It is low in the eastern part of our filament region because the photospheric fields are
more or less evenly spread over large areas. In the western part, there are strong compact sources of the field. They create a smaller scale field
in the corona that cannot be shaded by weaker larger scale fields. Cold plasma falls along both legs of an erupting prominence. It is observed in many events. The falling dark features are observed in the eastern leg of the  prominence too. When some parts of the erupting loop become nearly vertical, there is competition between the motion of the rising loop and falling back along the vertical flux tubes.

The distribution of the decay index shows that the instability zone  ($n >1$) appears at the position of the filament above 60 Mm (Fig. \ref{torus1}) . The instability zone spreads from the north-western side of the polarity inversion line (PIL) to the south-eastern side. We might expect that the north-western side of the filament is able to begin erupting first. However, this part of the filament is narrow and presumably low. The left panels of Fig. \ref{torus2} show PILs at different heights superposed on filtergrams containing the filament (the projection of the magnetic neutral surface on the filtergrams). The narrow north-western section of the filament is very close to low PILs (12 - 24 Mm). The opposite end of the filament is also rather narrow and low, while the middle section is widest and touches the PIL at the level of about 60 Mm. This is the height where we expect the instability of the flux rope, which started just before 17:00 UT. A day before, the filament was less wide in the middle section and less high.    This is in agreement with the observations showing that the middle part of the filament  started to rise first.

However, a question concerning the cause of the eruption is still open: How does the system come to the instability threshold?  It is commonly explained  by  photospheric  changes: horizontal motions, vertical motions (magnetic flux emergence and submergence) \citep{Schmieder2013},  canceling flux \citep{Martin1985}, or magnetic flux tube interactions \citep{Schmieder2004,Joshi2016}. Therefore we need to quantify the photospheric motions  by analyzing   the photospheric flows in the filament channel  and its environment. For that  we used the CST method.

 \section{Photospheric horizontal motions and structures}
 
 \begin{figure*} 
\centerline{\includegraphics[width=0.9\textwidth,clip=]{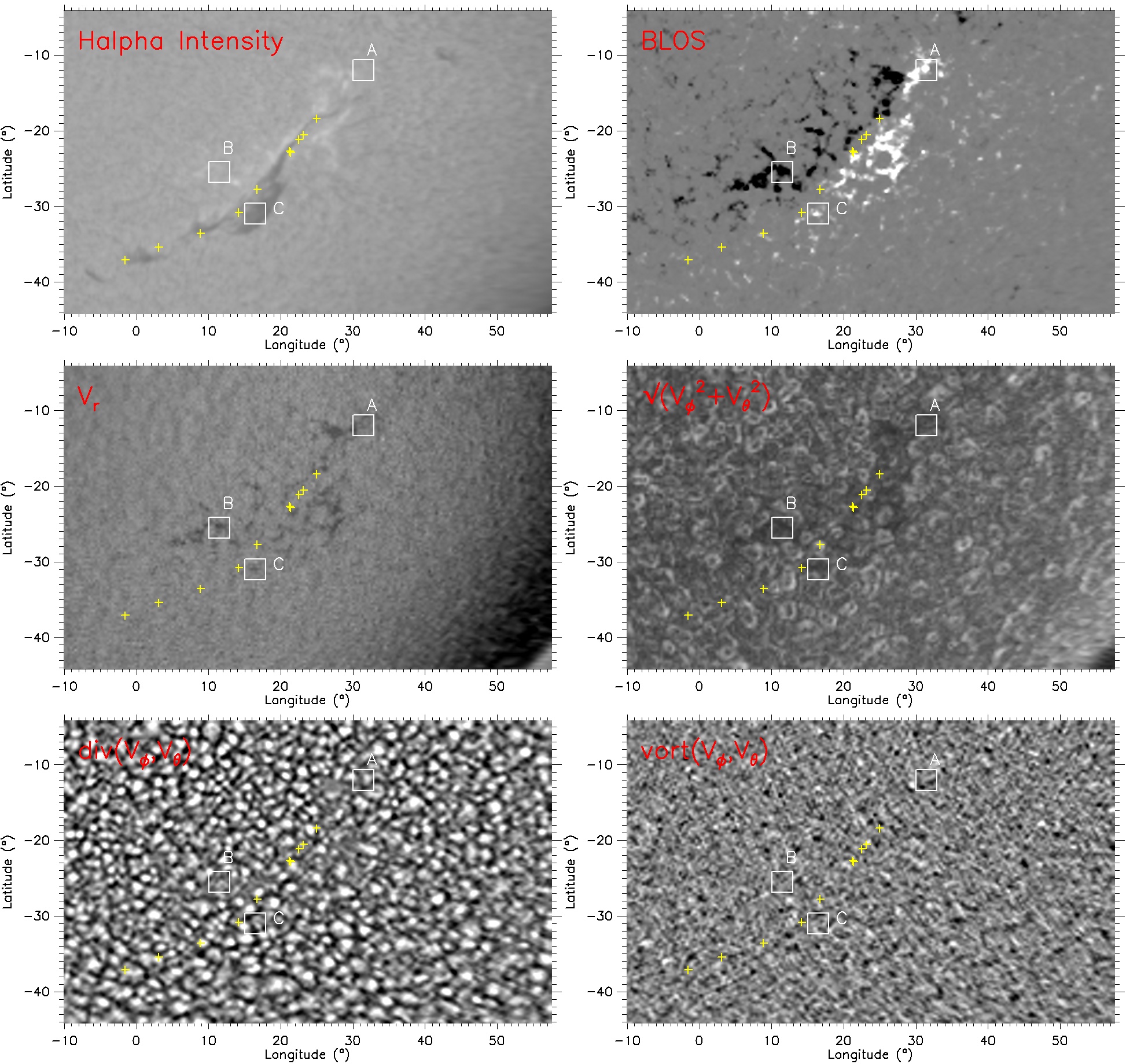}}
\caption{Top row left: H$\alpha$ filament  at 16:42 UT. The yellow crosses define  the polarity inversion line (PIL); the temporal evolution is available online Movie 1.  Top right:  HMI magnetogram
  (averaged over 24h). Middle row left: $ vr$ radial velocity (averaged over 24h),  white and dark indicating  outgoing (positive) and  ingoing (negative) radial velocity respectively, with the amplitude between 0.15 and -0.2 km/sec. Middle row right:   $vh$ horizontal velocity (averaged over 24h).
  Bottom row left: the  divergence (averaged over 24h) with a mean value of  ($ 3.6 10^{-4 }~s^{-1 }$) for positive divergence (white). Bottom row right : The vorticity (averaged over 24h),  with a mean value of  ($ 3.35 10^{-4 }~s^{-1 }$), for positive vorticity (white). 
 We note that the divergence is reduced in the filament channel around  the PIL  between A, B,  and C  (gray  areas) and is nearly co-spatial with  black  areas  in the $ vh$  modulus panel. In this map,  we see the supergranular pattern except in the regions  of black areas,  which correspond to strong magnetic field regions. For the four bottom panels, the granules were followed  during 24 hours to increase the contrast. The temporal evolution of the filament in $\alpha$ is available as a movie online. The movie shows the full disk, and the data in the movie before 16:00 UT are from CLIMSO and after 16:00 UT from GONG. CLIMSO and GONG data are centered and resized to the same solar diameter. The images in the movie are de-rotated from the solar differential rotation, allowing us to follow the filament evolution at the same location.
}
 \label{dia6}
 \end{figure*}

   \begin{figure*} 
 \centerline{
   \includegraphics[width=0.5\textwidth,clip=]{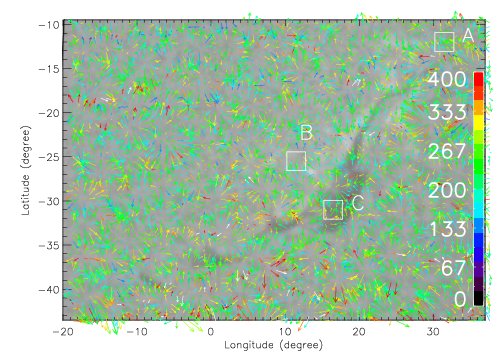}
  \includegraphics[width=0.5\textwidth,clip=]{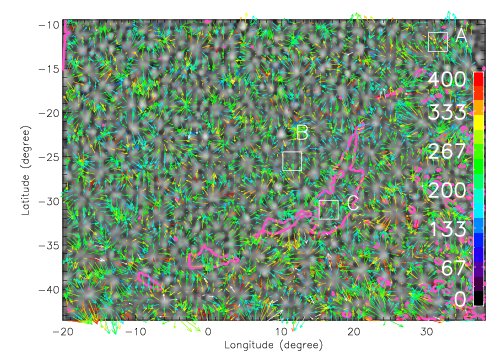}
   }
   \centerline{
  \includegraphics[width=0.5\textwidth,clip=]{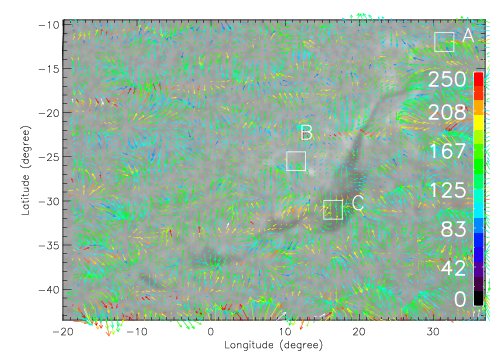} 
   \includegraphics[width=0.5\textwidth,clip=]{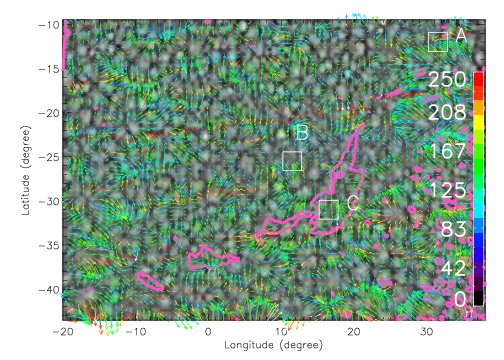}}
    \centerline{
    \includegraphics[width=0.5\textwidth,clip=]{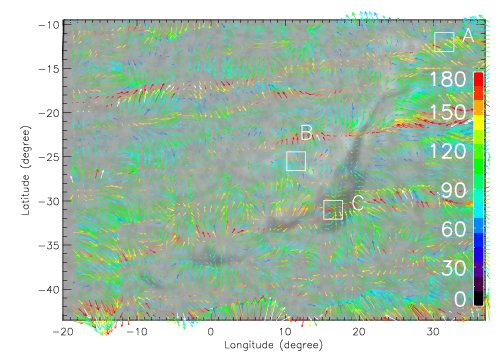} 
  \includegraphics[width=0.5\textwidth,clip=]{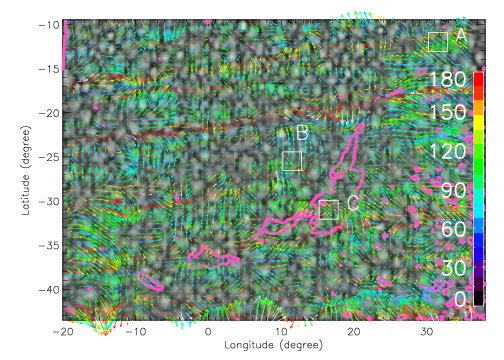}
 }
 \caption{Flows (computed between 09 and 16 UT) below the filament with three  different spatial  resolutions: (left panels) over an H$\alpha$ image (16:42 UT),  (right panels) over a divergence map. From top to bottom the spatial  window used is 22000 km - 44000 km - 66000 km, respectively. The granules have been followed during seven hours.
 The velocity is +/- 0.400, 0.250 km/s and  0.180  km/s for the  smaller to the largest windows.  A, B, and C are the ends of the filament (see Fig. \ref{SDO}). The amplitudes of the velocity  are  reduced in the filament channel  and are large on the west side of the filament.  We see three lanes of strong flows from the  west. The filament in H$\alpha$ is indicated by  pink  contours in the right panels.
  }
 \label{flow2}
 \end{figure*}

 \subsection{CST method}
 
 The representative description of solar plasma evolution is obtained by using the Coherent Structure Tracking \citep[CST][]{RRRD07,Roud2012} code to follow the proper motions of solar granules in full-disk SDO/HMI data. The CST allows us to obtain flow field from covering the spatial scales from 2.5 Mm up to nearly 85\% of the solar radius.  CST is a granule tracking technique that produces estimates of the field direction and amplitude \citep[][]{RRRD07}.  In CST, the code identifies individual features (granules) and tracks these individual features coherently throughout the image sequence. The resulting velocity field is thus discontinuous and the differentiable extension is estimated based on multi-resolution analysis.  
 HMI provides uninterrupted observations of the full disk of the Sun. This gives a unique opportunity to map surface flows on various spatial and temporal scales. We selected the HMI continuum intensity data on 26 January 2016.   In order to be suitable for the CST application, the data series of the HMI intensitygrams must first be prepared. All frames of the sequence are aligned such that the center of the solar disk lies exactly on the same pixel in
 charge-coupled device  (CCD) coordinates, and the radius of the solar disk is exactly the same for all the frames. The reference values for the position of the disk center and the radius are obtained from the first image  (obtained on 26 January 2016 at 00:00:45 UT). 
The original intensity and Doppler images are de-rotated by removing the differential rotation determined from 24 hours of Doppler data on 26 January 2016 (see next section) .

\subsection{Differential rotation}

It is well known that the profile of the differential rotation of the Sun differs by method, data set, and time. However, there have been rotational profiles published in the literature that are considered a reference. A class of the reference profiles was obtained by a spectroscopic method \citep[][] {Pat2010}, where historically the profile of,  for example, \citet[][]{HH70} is often used. 

Following the spectroscopic technique, we obtained our reference solar differential rotation profile from the corrected  sequence of Dopplergrams described above. That rotational profile is computed through the relation  
 
\begin{equation}
\Omega(\theta)=\frac{v_{\rm dop}(\theta,\phi)}{R \cos B_0 \cos\theta \sin\phi},
\end{equation}
where $\theta$ and $\phi$  are the latitude and longitude respectively, and $R$ the solar radius expressed in km. The profile inferred from an average 24hrs of observation is fitted by the polynomial in $sin(\theta)$, in microrad/s, by
  
\begin{eqnarray}
\Omega(\theta) &=& 2.86+0.00505 \sin\theta -0.549\sin^2\theta+\nonumber\\
                         &&+ 0.0015 \sin^3\theta -0.173 \sin^4\theta.
\end{eqnarray}The fit was performed in longitudes  $\pm 80\degr$ and the corresponding equatorial rotation velocity is $1.990\pm 0.002  $ km/s.

\subsection{Horizontal velocity}
Then we perform the granulation  tracking using the CST code \cite[][]{Roud2012} to reconstruct the projection of the photospheric velocity field ($v_{\rm x}$,$v_{\rm y}$) in the plane of the sky (CCD plane) \cite[][]{Rincon17}. The application of CST to  the data series of  the prepared HMI intensitygrams leads to a sequence of horizontal velocity field maps in the projection to a sky plane with a temporal resolution of 30 mins and a spatial resolution of 2.5 Mm (3\farcs5), which is the full-disk velocity map with a size of 586$\times$586~pixel$^2$. We further removed the $(x, y)$ velocity signal associated with the motions of the SDO satellite and Earth's orbital displacements from the CST velocity maps following the procedure described by \cite{Rincon17}. 

\subsection{Dopplergrams}
The dopplergrams provide a key piece of information to reconstruct the full vector field. The HMI convention is that the line-of-sight (l.o.s.) velocity signal $v_{\rm z}$ is taken to be positive when the flow is away from the observer, so that the out-of-plane toward the observer is $v_{\rm dop} =-v_{\rm z}$. \citep[see Fig. 10 in][]{RRPM13}. The processing steps of the Dopplergrams  were to remove Doppler  shift associated  with the proper motion of satellite and Earth displacement from the  raw Doppler signal \citep[see][]{Rincon17}. Then in the Doppler data we corrected a polynomial radial limbshift function adjusted from ring averages of two hours of data. Since the $586^2$~px$^2$  velocity maps obtained from the CST are limited to an effective resolution  of 2.5~Mm, we then down-sampled the $4096^2$~px$^2$ Dopplergrams to the size of CST maps. The down-sampled  Doppler images were finally averaged over 30 min to match the temporal sampling of the CST-derived flow maps.

\subsection{Velocity vector}

The horizontal flow $(v_x,v_y)$ measured in the plane-of-the-sky coordinates by the CST code together with  $v_{\rm dop}$ obtained from corresponding Dopplergrams may be transformed to spherical velocity components $(v_r, v_\phi, v_\theta)$. For a detailed description of the transformation,  we refer to \cite{RRPM13}, namely to Chapter 5 and Fig.~10. The  $v_\phi$ (longitudinal velocity) and  $v_{\theta} $ (latitudinal velocity) are used in the following as the  components of  the  horizontal  velocity vector ($vh$)  in the different plots.

 \section{Results}

 \subsection{Modulus of the  horizontal flow velocity and divergence}
 Using the HMI continuum, we applied the CST method to follow the granules over seven to 24 hours  with different spatial windows.
We computed the  modulus of   $v_\phi  $ and $ v_{\theta} $, always positive, to see the  global surface  motions:\\
 $vh=sqrt(v_\phi ^2 + v_\theta^2). $\\

The vector amplitude depends on the spatial window that we choose and the time that we use for following the granules.   The divergence and the  vorticity of the velocity  are  computed for  two different times of integration before the eruption (seven and 24 hours). Long integration increases the contrast of the  images.
Fig. \ref{dia6}  shows ,  in the middle the panels, the radial velocities $v_r$  and the modulus of the velocities ($vh$), and in the bottom panels the
divergence (div($v_\phi, v_\theta$))  and the vorticity (vort($v_\phi, v_\theta$)) after  following the granules during 24 h. We see in the divergence  map some areas of gray uniform color among the small bubbles that represent the supergranules. The low divergence areas  correspond to regions where the modulus of the vh velocity is reduced (dark areas in   the $vh$  velocity map).    They are co-spatial with the magnetized  areas of the   network (top right panel). These dark areas  of reduced vh  are located in the large filament channel including the ends of the filament in A, B, and C. The vorticity field does not appear affected by the magnetized areas (Fig. \ref{dia6}, bottom right). The radial velocities, perpendicular to the solar surface, are found to be negative (downflow motions) where the magnetic field is present around the PIL (Fig. \ref{dia6}, middle right).

\begin{figure} 
  \centerline{\includegraphics[width=0.5\textwidth,clip=]{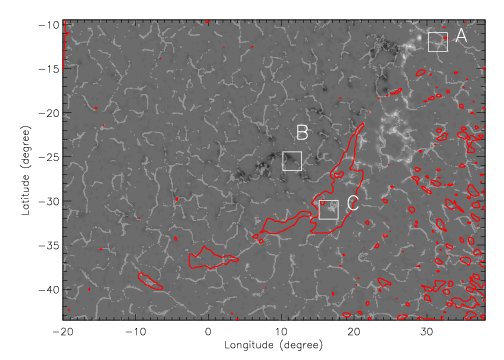}}
 \caption{Corks distribution after seven hours  tracking (from 09 to 16 UT),  visible as white lanes  overlaid over the magnetic field (17:00 UT). B is in the middle of the field of view  and corresponds to negative polarity (black), A is in the upper right  corner in the positive polarity (white). C is in the white polarity.  The filament in H$\alpha$ is indicated by  red  contours.}
 \label{cork}
 \end{figure}

Figure \ref{flow2}  presents  the map of  the vector velocity: (right panels)  overlying  the divergence map and (left panels) overlying an H$\alpha$ image for
three different windows (from top to bottom: 22 000 km, 44 000 km, 66 000 km).   The integration time is seven hours. We note three horizontal lines of higher velocities, which cross the filament channel when we use  the largest window.
However, we see that  the velocities are always  reduced in the filament channel and  also in the  divergence  compared to   both side areas  of  the filament channel.  This behavior does not depend on the size of the  chosen windows. Figure \ref{cork} presents the concentration of the corks  (white lanes)  after launching them uniformly seven hours before. Corks are diffused over the solar surface by the horizontal flows. They are concentrated over the network where the ends of the filament are anchored. Convergence lanes are in the vicinity of  points A,  B, and C (Fig. \ref{cork}).  The $vh$  velocity modulus is found up to 0.400 km/s, 0.250 km/s  for narrow and middle size windows compared with 0.180 km/s for the largest  windows. We measure downflows along the PIL  ($v_r$  around 0.100 to 0.200 km/s) indicating that the magnetic field is well anchored in the deeper layers and probably stable.

\subsection{Local horizontal flows} 

\begin{figure} 
    \centerline{
\includegraphics[width=0.5\textwidth,clip=]{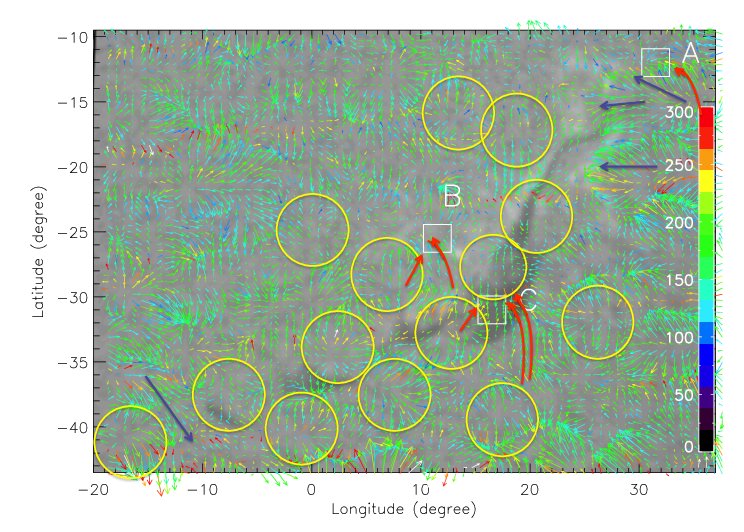}}  
\caption{ Horizontal velocities (computed between 09 and 16 UT) over an H$\alpha$ image (16:42 UT).
Yellow contours  indicate the  supergranules. The black arrows indicate flows going through the filament channel  from the north towards the south in the eastern part or from west to east in the northern part,  perpendicular to  the filament axis. The  red arrows indicate  flows perpendicular to the filament with some whirls.     The large west- east flows on the western  side of the filament are stopped at the western edge of the filament channel.
}
 \label{dia9}
 \end{figure}

 In Fig. \ref{dia9} we have drawn the contours of a few supergranules and emphasized the large flows by arrows over the horizontal velocity flows obtained with a window of 44 000 km  presented already  in Fig. \ref{flow2}. The     supergranular pattern is well observed  
  in  and outside the filament channel with no real changes of size.
 The black arrows represent flows that cross the filament channel. The red arrows represent the flows that are stopped at the border of  the filament channel. 

 The  velocity  vector maps  show   large flows   on the west side (right side) of the filament channel. They    are stopped   close to  the channel  and turned to the north (near points A, B, and C). Below the large extension of the filament, where the end of the south section of  Filament two (point C)  is anchored, high speed flows  from  a SW to NE direction meet with other flows coming from the NE.  These flows stop in C.  We could guess that they initiate vertical motions and could be  the trigger for the rise of the filament  that allows the filament to reach the altitude where the torus instability can operate. 
In  the eastern part  of the filament,  the flow  runs from north to  south, crossing the filament channel.  In  B region, which corresponds to  a strong network magnetic field, large flows converge in the region  and push the magnetic field lines,  which are sliding along the magnetized region. During the eruption (see Movie 2  of Fig. 1, left panel) Whirled flows are found locally around  A, B, and C.

\subsection{Differential rotation and shear flow}

\begin{figure} 
 \centerline{
 \includegraphics[width=0.5\textwidth,clip=]{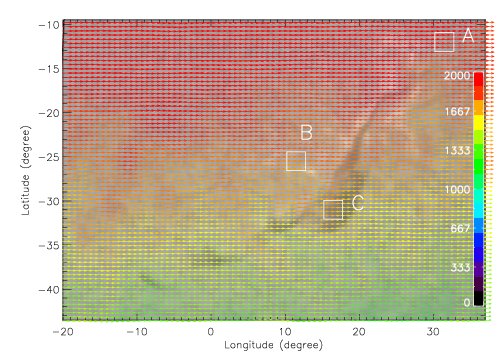}}
 \caption{  Horizontal flows (spherical velocity components $(v_\phi, v_\theta)$ ) computed between 9 and 16 UT including the differential rotation under  the filament observed at 16:42 UT.  The main flow is the solar rotation vector from right to left disrupted by other surface flows (mainly supergranules). The velocity is reduced on the western part of the filament compared to the eastern part. It indicates  the presence of convergence along the filament PIL.
  }
 \label{rot}
 \end{figure}

The velocity in the preceding sections  has been measured on de-rotated data. In order to take into account the large flow undergone by the
filament feet, we now add the solar differential rotation that was measured on the Doppler data (see Sect. 3.2). 
Figure \ref{rot} presents the horizontal flows (spherical velocity components $(v_\phi, v_\theta)$) including the differential rotation.
 That plot shows the full motions on the solar surface, which are a combination of the solar rotation and other larger flows provided by the conjugation of all the supergranular flows. Stronger velocity amplitude is found close to the equator with a decrease at higher latitude.  It shows strong flows from west to east reducing  the differential rotation velocity in the central part of the filament (B and C) where    the anchorages of
the field lines of the flux rope are sustaining the filament. At point A, the velocity, mainly due to solar rotation, is found to be 1.82 km/s. At point C, the velocity is measured at 1.53 km/s
while at this latitude ($31\degr$ south) that day, the average velocity is 1.63 km/s. This difference increases the shearing experienced by the filament
between points A and C. We also note, below point C  at longitude $8\degr$ and latitude $-37\degr$  , some larger velocities
(1.48 km/s , while solar rotation at that latitude is 1.35 km/s. These velocities are located between Filament one and  Filament two and are probably involved in the filament eruption. The  PIL corresponding to the filament  in the photosphere is  submitted to strong convergence flows creating a strong shear flow.

 \section{Conclusion}

  We analyzed the dynamics of  the photosphere below a  long filament during 24 hours on January 26 2016  before its eruption. 
Counterstreamings along the filament before its eruption  are observed in   H$\alpha$  , as well as  moving brightenings in 304 \AA\ along its spine
during all this time period. The main results  of the Coherent Structure Tracking (CST)  analysis concerning the flows in the photosphere are the following:\\
\begin{itemize}
\item  Supergranules  with diverging flow pattern have a similar  size in and outside the filament channel but have a lower amplitude in the magnetized
  areas. In the same way,  the modulus of the horizontal velocity is reduced in the filament channel, particularly in the magnetized region where
  the field lines of the flux rope of the filament  are anchored (A, B, C)
\item  The  ends of the two filament sections (A, B, C) are between supergranules in regions of convergent flows nearly perpendicular to the axis
  of the filament.
\item Whirled flows are found locally around points A, B, and C.
\item  Corks meet  in  magnetized areas (convergence areas)   and  consequently are associated to the filament ends.  
\item  Large zones of flows lie at the  western border of the filament channel according to the analysis of  large-scale coherent structures.
\item Strong flows from west to east reduce the differential rotation velocity in the central part of the filament where  the anchorage of the field lines are in B and C. That difference in flows creates a stronger shear flow experienced by the filament feets. 
\end{itemize}

There is a strong coupling between convection and magnetic field in the photosphere.  Large convection cells transport the magnetic field and  form between them a magnetic polarity inversion line. We have seen convergence of the flows towards the inversion line PIL.  We conjectured that the different loops  or arcades  over the magnetic PIL are sheared and reconnection is possible due the convergence motions  like in the model of {\citet{vanBallegooijen1989}. Through the reconnections,  a  twisted flux rope is formed  like in  the simulation \citep{Aulanier2010}.  It  could  correspond to   the filament that we  have observed. If the convergence motion continues, then  the convection cells are continuing to roll  one against to each other; material in the photosphere is going down  along the PIL. We have effectively measured  downflow velocities  between - 0.100 and  - 0.200 km/s while the filament was rising. When it reaches the height of  the torus instability threshold, it  could erupt. The horizontal photospheric motions  that we have measured explain the formation and the eruption of the filament. The CST method could be also generalized and applied systematically to full disk HMI magnetograms.


  \begin{acknowledgements}
    
The SDO data are courtesy of NASA and the SDO/HMI, SDO/AIA, and SOHO science team.  The CLIMSO instrument  and  its  operation are
funded  respectively  by  the  Fiducial  company  and  by  French  government  funds through University Paul Sabatier and Centre National de la
Recherche Scientifique (CNRS). This work utilizes data obtained by the Global Oscillation Network Group (GONG) program, managed by the National
Solar Observatory, which is operated by AURA, Inc. under a cooperative agreement with the National Science Foundation. The data were acquired by
instruments operated by the Big Bear Solar Observatory, High Altitude Observatory, Learmonth Solar Observatory, Udaipur Solar Observatory, Instituto de
Astrof\'{\i}sica de Canarias, and Cerro Tololo Interamerican Observatory. This work was granted access to the HPC resources of CALMIP
under the allocation 2011-[P1115]. This work was supported by the CNRS Programme National Soleil Terre. This work was inspired during discussions in the ISSI workshops organized by Nicolas Labrosse in Berne on  {\it Solving the prominence paradox}.  RC acknowledges the support from SERB/DST, New Delhi,
  Govt. of India project no. SERB/F/7455/2017-17. We thank the anonymous referee for his/her careful reading of our manuscript and his/her many insightful comments and suggestions.

\end{acknowledgements}

\bibliographystyle{aa}    
\bibliography{roudier_bib}

\begin{thebibliography}{53}
\expandafter\ifx\csname natexlab\endcsname\relax\def\natexlab#1{#1}\fi

\bibitem[{{Ahn} {et~al.}(2010){Ahn}, {Chae}, {Cao}, \& {Goode}}]{Ahn10}
{Ahn}, K., {Chae}, J., {Cao}, W., \& {Goode}, P.~R. 2010, \apj, 721, 74

\bibitem[{{Aulanier} \& {D\'emoulin}(1998)}]{Aulanier98}
{Aulanier}, G. \& {D\'emoulin}, P. 1998, \aap, 329, 1125

\bibitem[{{Aulanier} \& {Schmieder}(2002)}]{Aulanier2002}
{Aulanier}, G. \& {Schmieder}, B. 2002, \aap, 386, 1106

\bibitem[{{Aulanier} {et~al.}(2010){Aulanier}, {T{\"o}r{\"o}k}, {D{\'e}moulin},
  \& {DeLuca}}]{Aulanier2010}
{Aulanier}, G., {T{\"o}r{\"o}k}, T., {D{\'e}moulin}, P., \& {DeLuca}, E.~E.
  2010, \apj, 708, 314

\bibitem[{{Chae} \& {Sakurai}(2008)}]{Chae08}
{Chae}, J. \& {Sakurai}, T. 2008, \apj, 689, 593

\bibitem[{{Chian} {et~al.}(2014){Chian}, {Rempel}, {Aulanier}, {Schmieder},
  {Shadden}, {Welsch}, \& {Yeates}}]{Chian2014}
{Chian}, A.~C.-L., {Rempel}, E.~L., {Aulanier}, G., {et~al.} 2014, \apj, 786,
  51

\bibitem[{{D{\'e}moulin} \& {Aulanier}(2010)}]{Aula2010}
{D{\'e}moulin}, P. \& {Aulanier}, G. 2010, \apj, 718, 1388

\bibitem[{{Filippov} \& {Zagnetko}(2008)}]{Filippov2008}
{Filippov}, B. \& {Zagnetko}, A. 2008, Journal of Atmospheric and
  Solar-Terrestrial Physics, 70, 614

\bibitem[{{Filippov} \& {Den}(2001)}]{Filippov2001}
{Filippov}, B.~P. \& {Den}, O.~G. 2001, \jgr, 106, 25177

\bibitem[{{Green} {et~al.}(2011){Green}, {Kliem}, \& {Wallace}}]{Green11}
{Green}, L.~M., {Kliem}, B., \& {Wallace}, A.~J. 2011, \aap, 526, A2

\bibitem[{{Hermans} \& {Martin}(1986)}]{Hermans1986}
{Hermans}, L.~M. \& {Martin}, S.~F. 1986, in NASA Conference Publication, Vol.
  2442, NASA Conference Publication, ed. A.~I. {Poland}

\bibitem[{{Howard} \& {Harvey}(1970)}]{HH70}
{Howard}, R. \& {Harvey}, J. 1970, \solphys, 12, 23

\bibitem[{{Joshi} {et~al.}(2016){Joshi}, {Filippov}, {Schmieder}, {Magara},
  {Moon}, \& {Uddin}}]{Joshi2016}
{Joshi}, N.~C., {Filippov}, B., {Schmieder}, B., {et~al.} 2016, \apj, 825, 123

\bibitem[{{Kliem} {et~al.}(2014){Kliem}, {Lin}, {Forbes}, {Priest}, \&
  {T{\"o}r{\"o}k}}]{Kliem2014}
{Kliem}, B., {Lin}, J., {Forbes}, T.~G., {Priest}, E.~R., \& {T{\"o}r{\"o}k},
  T. 2014, \apj, 789, 46

\bibitem[{{Kliem} \& {T{\"o}r{\"o}k}(2006)}]{Kliem2006}
{Kliem}, B. \& {T{\"o}r{\"o}k}, T. 2006, Physical Review Letters, 96, 255002

\bibitem[{{Lemen} {et~al.}(2012){Lemen}, {Title}, {Akin}, {Boerner}, {Chou},
  {Drake}, {Duncan}, {Edwards}, {Friedlaender}, {Heyman}, {Hurlburt}, {Katz},
  {Kushner}, {Levay}, {Lindgren}, {Mathur}, {McFeaters}, {Mitchell}, {Rehse},
  {Schrijver}, {Springer}, {Stern}, {Tarbell}, {Wuelser}, {Wolfson}, {Yanari},
  {Bookbinder}, {Cheimets}, {Caldwell}, {Deluca}, {Gates}, {Golub}, {Park},
  {Podgorski}, {Bush}, {Scherrer}, {Gummin}, {Smith}, {Auker}, {Jerram},
  {Pool}, {Soufli}, {Windt}, {Beardsley}, {Clapp}, {Lang}, \&
  {Waltham}}]{Lemen2012}
{Lemen}, J.~R., {Title}, A.~M., {Akin}, D.~J., {et~al.} 2012, \solphys, 275, 17

\bibitem[{{Litvinenko} \& {Martin}(1999)}]{Litvinenko1999}
{Litvinenko}, Y.~E. \& {Martin}, S.~F. 1999, \solphys, 190, 45

\bibitem[{{Liu} {et~al.}(2012){Liu}, {Berger}, \& {Low}}]{Liu12}
{Liu}, W., {Berger}, T.~E., \& {Low}, B.~C. 2012, \apjl, 745, L21

\bibitem[{{Luna} {et~al.}(2017){Luna}, {Su}, {Schmieder}, {Chandra}, \&
  {Kucera}}]{Luna2017}
{Luna}, M., {Su}, Y., {Schmieder}, B., {Chandra}, R., \& {Kucera}, T.~A. 2017,
  \apj, 850, 143

\bibitem[{{Martin}(1986)}]{Martin1986}
{Martin}, S.~F. 1986, in NASA Conference Publication, Vol. 2442, NASA
  Conference Publication, ed. A.~I. {Poland}

\bibitem[{{Martin}(1998)}]{Martin1998}
{Martin}, S.~F. 1998, in Astronomical Society of the Pacific Conference Series,
  Vol. 150, IAU Colloq. 167: New Perspectives on Solar Prominences, ed. D.~F.
  {Webb}, B.~{Schmieder}, \& D.~M. {Rust}, 419

\bibitem[{{Martin} {et~al.}(1994){Martin}, {Bilimoria}, \&
  {Tracadas}}]{Martin94}
{Martin}, S.~F., {Bilimoria}, R., \& {Tracadas}, P.~W. 1994, in Solar Surface
  Magnetism, ed. R.~J. {Rutten} \& C.~J. {Schrijver}, 303

\bibitem[{{Martin} {et~al.}(1985){Martin}, {Livi}, \& {Wang}}]{Martin1985}
{Martin}, S.~F., {Livi}, S.~H.~B., \& {Wang}, J. 1985, Australian Journal of
  Physics, 38, 929

\bibitem[{{November} \& {Simon}(1988)}]{November88}
{November}, L.~J. \& {Simon}, G.~W. 1988, \apj, 333, 427

\bibitem[{{Patern{\`o}}(2010)}]{Pat2010}
{Patern{\`o}}, L. 2010, \apss, 328, 269

\bibitem[{{Pesnell} {et~al.}(2012){Pesnell}, {Thompson}, \&
  {Chamberlin}}]{Pesnell2012}
{Pesnell}, W.~D., {Thompson}, B.~J., \& {Chamberlin}, P.~C. 2012, \solphys,
  275, 3

\bibitem[{{Priest}(1987)}]{Priest1987}
{Priest}, E.~R. 1987, in The Role of Fine-Scale Magnetic Fields on the
  Structure of the Solar Atmosphere, ed. E.-H. {Schr{\"o}ter},
  M.~{V{\'a}zquez}, \& A.~A. {Wyller}, 297

\bibitem[{{Priest} {et~al.}(1994){Priest}, {Parnell}, \& {Martin}}]{Priest1994}
{Priest}, E.~R., {Parnell}, C.~E., \& {Martin}, S.~F. 1994, \apj, 427, 459

\bibitem[{Rieutord {et~al.}(2007)Rieutord, Roudier, Roques, \&
  Ducottet}]{RRRD07}
Rieutord, M., Roudier, T., Roques, S., \& Ducottet, C. 2007, \aap, 471, 687

\bibitem[{{Rincon} {et~al.}(2017){Rincon}, {Roudier}, {Schekochihin}, \&
  {Rieutord}}]{Rincon17}
{Rincon}, F., {Roudier}, T., {Schekochihin}, A.~A., \& {Rieutord}, M. 2017,
  \aap, 599, A69

\bibitem[{{Rondi} {et~al.}(2007){Rondi}, {Roudier}, {Molodij}, {Bommier},
  {Keil}, {S{\"u}tterlin}, {Malherbe}, {Meunier}, {Schmieder}, \&
  {Maloney}}]{Rondi07}
{Rondi}, S., {Roudier}, T., {Molodij}, G., {et~al.} 2007, \aap, 467, 1289

\bibitem[{{Roudier} {et~al.}(2012){Roudier}, {Rieutord}, {Malherbe}, {Renon},
  {Berger}, {Frank}, {Prat}, {Gizon}, \& {{\v S}vanda}}]{Roud2012}
{Roudier}, T., {Rieutord}, M., {Malherbe}, J.~M., {et~al.} 2012, \aap, 540, A88

\bibitem[{{Roudier} {et~al.}(2013){Roudier}, {Rieutord}, {Prat}, {Malherbe},
  {Renon}, {Frank}, {{\v S}vanda}, {Berger}, {Burston}, \& {Gizon}}]{RRPM13}
{Roudier}, T., {Rieutord}, M., {Prat}, V., {et~al.} 2013, \aap, 552, A113

\bibitem[{{Roudier} {et~al.}(2008){Roudier}, {{\v S}vanda}, {Meunier}, {Keil},
  {Rieutord}, {Malherbe}, {Rondi}, {Molodij}, {Bommier}, \&
  {Schmieder}}]{Roudier08}
{Roudier}, T., {{\v S}vanda}, M., {Meunier}, N., {et~al.} 2008, \aap, 480, 255

\bibitem[{{Roudier} {et~al.}(2018){Roudier}, {{\v{S}}vanda}, {Ballot},
  {Malherbe}, \& {Rieutord}}]{Roudier2018}
{Roudier}, T., {{\v{S}}vanda}, M., {Ballot}, J., {Malherbe}, J.~M., \&
  {Rieutord}, M. 2018, \aap, 611, A92

\bibitem[{{Scherrer} {et~al.}(2012){Scherrer}, {Schou}, {Bush}, {Kosovichev},
  {Bogart}, {Hoeksema}, {Liu}, {Duvall}, {Zhao}, {Title}, {Schrijver},
  {Tarbell}, \& {Tomczyk}}]{Scherrer2012}
{Scherrer}, P.~H., {Schou}, J., {Bush}, R.~I., {et~al.} 2012, \solphys, 275,
  207

\bibitem[{{Schmieder} {et~al.}(2006){Schmieder}, {Aulanier}, {Mein}, \&
  {L{\'o}pez Ariste}}]{Schmieder06}
{Schmieder}, B., {Aulanier}, G., {Mein}, P., \& {L{\'o}pez Ariste}, A. 2006,
  \solphys, 238, 245

\bibitem[{{Schmieder} {et~al.}(2008){Schmieder}, {Bommier}, {Kitai},
  {Matsumoto}, {Ishii}, {Hagino}, {Li}, \& {Golub}}]{Schmieder08}
{Schmieder}, B., {Bommier}, V., {Kitai}, R., {et~al.} 2008, \solphys, 247, 321

\bibitem[{{Schmieder} {et~al.}(2013){Schmieder}, {D{\'e}moulin}, \&
  {Aulanier}}]{Schmieder2013}
{Schmieder}, B., {D{\'e}moulin}, P., \& {Aulanier}, G. 2013, Advances in Space
  Research, 51, 1967

\bibitem[{{Schmieder} {et~al.}(2004){Schmieder}, {Mein}, {Deng}, {Dumitrache},
  {Malherbe}, {Staiger}, \& {Deluca}}]{Schmieder2004}
{Schmieder}, B., {Mein}, N., {Deng}, Y., {et~al.} 2004, \solphys, 223, 119

\bibitem[{{Schmieder} {et~al.}(2014){Schmieder}, {Roudier}, {Mein}, {Mein},
  {Malherbe}, \& {Chandra}}]{Schmieder2014}
{Schmieder}, B., {Roudier}, T., {Mein}, N., {et~al.} 2014, \aap, 564, A104

\bibitem[{{Schou} {et~al.}(2012){Schou}, {Scherrer}, {Bush}, {Wachter},
  {Couvidat}, {Rabello-Soares}, {Bogart}, {Hoeksema}, {Liu}, {Duvall}, {Akin},
  {Allard}, {Miles}, {Rairden}, {Shine}, {Tarbell}, {Title}, {Wolfson},
  {Elmore}, {Norton}, \& {Tomczyk}}]{Schou2012}
{Schou}, J., {Scherrer}, P.~H., {Bush}, R.~I., {et~al.} 2012, \solphys, 275,
  229

\bibitem[{{Schuck}(2008)}]{Schuck08}
{Schuck}, P.~W. 2008, \apj, 683, 1134

\bibitem[{{T{\"o}r{\"o}k} \& {Kliem}(2007)}]{Torok2007}
{T{\"o}r{\"o}k}, T. \& {Kliem}, B. 2007, Astronomische Nachrichten, 328, 743

\bibitem[{{van Ballegooijen} \& {Martens}(1989)}]{vanBallegooijen1989}
{van Ballegooijen}, A.~A. \& {Martens}, P.~C.~H. 1989, \apj, 343, 971

\bibitem[{{van Tend} \& {Kuperus}(1978)}]{Tend1978}
{van Tend}, W. \& {Kuperus}, M. 1978, \solphys, 59, 115

\bibitem[{{Wang} \& {Muglach}(2013)}]{Wang2013}
{Wang}, Y.-M. \& {Muglach}, K. 2013, \apj, 763, 97

\bibitem[{{Welsch} {et~al.}(2007){Welsch}, {Abbett}, {De Rosa}, {Fisher},
  {Georgoulis}, {Kusano}, {Longcope}, {Ravindra}, \& {Schuck}}]{Welsch2007}
{Welsch}, B.~T., {Abbett}, W.~P., {De Rosa}, M.~L., {et~al.} 2007, \apj, 670,
  1434

\bibitem[{{Yardley} {et~al.}(2016){Yardley}, {Green}, {Williams}, {van
  Driel-Gesztelyi}, {Valori}, \& {Dacie}}]{Yardley2016}
{Yardley}, S.~L., {Green}, L.~M., {Williams}, D.~R., {et~al.} 2016, \apj, 827,
  151

\bibitem[{{Zheng} {et~al.}(2017){Zheng}, {Zhang}, {Chen}, {Wang}, {Du}, {Li},
  \& {Yang}}]{Zheng2017}
{Zheng}, R., {Zhang}, Q., {Chen}, Y., {et~al.} 2017, \apj, 836, 160

\bibitem[{{Zhou} {et~al.}(2006){Zhou}, {Wang}, \& {Wang}}]{Zhou06}
{Zhou}, G., {Wang}, Y., \& {Wang}, J. 2006, Advances in Space Research, 38, 466

\bibitem[{{Zirker} {et~al.}(1998){Zirker}, {Engvold}, \& {Martin}}]{Zirker98}
{Zirker}, J.~B., {Engvold}, O., \& {Martin}, S.~F. 1998, \nat, 396, 440

\bibitem[{{Zuccarello} {et~al.}(2016){Zuccarello}, {Aulanier}, \&
  {Gilchrist}}]{Zuccarello2016}
{Zuccarello}, F.~P., {Aulanier}, G., \& {Gilchrist}, S.~A. 2016, \apjl, 821,
  L23

\end{thebibliography}

\end{document}